\documentclass[aps,10pt,nofootinbib,preprint]{revtex4-1}
\usepackage[T1]{fontenc}
\usepackage{hyperref}
\usepackage{xcolor}
\usepackage{graphicx}
\usepackage{amsmath,amssymb}
\usepackage{bm}

\begin{document}
	
	\title{\boldmath Conformal frames in cosmology}
	
	\author{{Guillem Dom\`enech$^*$} and {Misao Sasaki$^\dag$}}
	\address{Yukawa Institute for Theoretical Physics,\\ 
		Kyoto University, Kyoto 606-8502, Japan\\
		$^*$guillem.domenech{}@{}yukawa.kyoto-u.ac.jp\\
		$\dag$misao{}@{}yukawa.kyoto-u.ac.jp}
	\date{\today}
	\preprint{YITP-16-11}

	\begin{abstract}
    From higher dimensional theories, e.g. string theory, one expects the presence of non-minimally coupled scalar fields. We review the notion of conformal frames in cosmology and emphasize their physical equivalence, which holds at least at a classical level.
    Furthermore, if there is a field, or fields, which dominates the universe, as it is often the case in cosmology, we can use such notion of frames to treat our system, matter and gravity, as two different sectors. On one hand, the gravity sector which describes the dynamics of the geometry and on the other hand the matter sector which has such geometry as a playground. We use this interpretation to build a model where the fact that a curvaton couples to a particular frame metric could leave an imprint in the CMB.\\\\\\\\\\\\\\\\\\\\\\
    Note: Prepared for the Proceedings for the 2nd LeCosPA Symposium:
    Everything about Gravity.
	\end{abstract}
	
\maketitle
	
	\section{Introduction: Conformal frames \label{intro}}
	
	In cosmology, it is certainly often assumed that some field(s) rules the evolution of the universe, i.e. dictates metric dynamics, while the rest, say matter, plays out its dynamics in such a given geometry. There is no reason, at least on cosmological scales, to a priori assume that the playground geometry for matter, from now on we call it \textit{matter metric}, is exactly equal to that given by the dominating fields, for example if those fields are non-minimally coupled to gravity. These metrics could be non-trivially related through a function of the fields, the simplest case being the so-called conformal transformation, that is a rescaling of the metric. Needless to say, these two geometries must be almost indistinguishable on solar system scales, where very strong constraints apply.
	
	One may wonder if such model, i.e. two different geometries, could be derived from a theoretical model rather than be postulated. In fact, without going into much details, it is fairly easy to see that in higher dimensional spacetimes this is not surprising at all\cite{fujii2003scalar}. For example, consider a D-dimensional space where the metric is given by
	\begin{align}
		\left(
		\begin{array}{cc}
			g^{(D)}_{\mu\nu}(x,y)& g^{(D)}_{\mu B}(x,y)\\
			g^{(D)}_{A\nu}(x,y)& g^{(D)}_{AB}(x,y)
		\end{array}	
		\right)
	\end{align}
	where $\mu,\nu$=1...4, $A,B$=4...D and $x$ and $y$ are respectively 4 and 4-D dimensional coordinates. As matter is concerned, we know it lives in a 4 dimensional spacetime. How is the dimensionality reduced or, in other words, which is the effective metric for matter, covers many possibilities. To gather some examples, one could have:
			\begin{align}
						g_{\mu\nu}(x)\stackrel{?}{=}
				\left\{
				\begin{array}{c}
					\langle g^{(D)}_{\mu\nu}\rangle_{4-D}\\
					f(x)\langle g^{(D)}_{\mu\nu}\rangle_{4-D}\\
					g^{(D)}_{\mu\nu}(x,0)\\
					...
				\end{array}	
				\right.\,\,,
			\end{align}
	where brackets mean that we integrated out the 4-D extra dimensions. Under a dimensional reduction, the initial D tensor modes translate into 4 tensor, 4 vector and 4 scalar modes. Most likely, such dimensional reduction ends up involving dilatonic scalars, that is scalars fields which non-minimally couple to matter and/or gravity. In that sense, it is clear that there is no unique natural conformal or physical frame a priori.
	
	For concreteness sake, let us show two typical conformal frames in cosmology and let us work in planck units ($\hbar=c=M_{pl}=1$). On one hand, the Jordan frame\cite{jordan1959gegenwartigen,brans1961mach} where the action is given by
	\begin{align}
		S=\int d^4x \sqrt{-\bar g}\left\{F(\phi)\bar R+{\cal L}(\phi)+{\cal L}_{m}(\psi,A,...)\right\}\,,
	\end{align}
	where matter fields, e.g. fermions $\psi$ and vectors $A$, minimally couple to the metric $\bar g$ but there is a non-minimal coupling between a scalar field $\phi$ and gravity. On the other hand the Einstein frame action is
	\begin{align}
		S=\int d^4x \sqrt{-g}\left\{R+{\cal L}(\phi)+G(\phi){\cal L}_{m}(\phi,\psi,A,...)\right\}\,,
	\end{align}
	where this time the scalar field $\phi$ minimally couples to the metric $g$ but there appears a coupling between matter fields and the scalar field $\phi$. In the Jordan frame, one usually assumes that matter is \textit{universally} coupled, which for baryons is experimentally consistent. If there were a non-universal coupling, we could write the matter Lagrangian as
	\begin{align}
	\sum_A G_A(\phi){\cal L}_{A}(\phi,Q_A)\,,
	\end{align}
	where $Q_A=\{\psi,A,...\}$. However, it should be noted that in the latter case the definition of Jordan frame, i.e. where matter is minimally coupled, is not clear and perhaps one could define a Jordan frame for each $Q_A$, if any.
	
	A note is in order. Throughout the foregoing discussion we did not emphasise any frame as more physical than others and, furthermore, from our dimensional reduction point of view there is no reason to believe that such a frame exist. 
	In next section, we review some examples of how physics does not change but interpretations do. Before that, let us remind the reader below how quantities transform under a conformal transformation as we will make use of them.
	\paragraph{Transformation rules:}
	Under a conformal transformation given by
	\begin{align}\label{eq:conf}
		\tilde g_{\mu\nu}=\Omega^2 g_{\mu\nu}
	\end{align}
	the Ricci scalar transforms as
	\begin{align}
		\tilde R=\Omega^{-2}\left[R-(D-1)\left(2\frac{\Box\Omega}{\Omega}-(D-4)g^{\mu\nu}\frac{\nabla_\mu\Omega\nabla_\nu\Omega}{\Omega^2}\right)\right]
	\end{align}
	and if one can neglect the dynamics of the dilaton field at shorts scales one can find that matter fields transform as
	\begin{align}
		\begin{array}{cc}\medskip
		\tilde \chi=\Omega^{-(D-2)/2}\chi&\qquad {\rm for~scalars\,,}\\\medskip
		\tilde A_\mu=\Omega^{-(D-4)/2}A_\mu&\qquad {\rm for~vectors\,,}\\
		\tilde \psi=\Omega^{-(D-1)/2}\psi &\qquad {\rm for~fermions\,.}
		\end{array}
	\end{align}
	
	Let us show in more detail the fermion and gauge field cases, as we use this results below. The action for a Dirac fermion is given by
	\begin{align}
		S=\int d^4x \sqrt{-g}\left[-i\bar\psi\gamma^\mu D_\mu\psi-m\bar\psi\psi-\frac{1}{4}F_{\mu\nu}F^{\mu\nu}\right]
	\end{align}
	where $D_\mu=\partial_\mu+ieA_\mu-\frac{1}{4}\omega_{ab\mu}\Sigma^{ab}$, $\gamma^\mu$ are the gamma matrices,  $\Sigma^{ab}=\frac{1}{2}\left[\gamma^a,\gamma^b\right]$ and $\omega_{ab\mu}=e_{a\nu}\nabla_{\mu}e^\nu_b$, $m$ is the mass of the fermion and $F_{\mu\nu}=\nabla_\mu A_\nu-\nabla_\mu A_\nu$.
	After the conformal transformation \eqref{eq:conf} one obtains
	\begin{align}\label{eq:fermion}
		S=\int d^4x \sqrt{-\tilde g}\left[-i\bar{\tilde\psi}\tilde\gamma^\mu D_\mu\tilde\psi-\mathbf{\tilde m}\bar{\tilde\psi}\tilde\psi-\frac{1}{4}F_{\mu\nu}F^{\mu\nu}\right]
	\end{align}
	where $\tilde\gamma^\mu=\Omega^{-1}\gamma^\mu$, $\tilde\psi=\Omega^{-3/2}\psi$, $\tilde m=\Omega^{-1} m$ and the gauge field $A_\mu$ is conformal invariant in 4 dimensions. We highlighted the mass in bold as it is the main effect of a conformal transformation; the mass of the fermion becomes spacetime dependent.
	
	\section{Frame independence of observables \label{frames}}
	The physical equivalence of conformal frames has been extensively discussed in the literature\cite{makino1991density, faraoni2007pseudo, deruelle2011conformal, gong2011conformal, white2012curvature, jarv2014invariant,Kuusk:2015dda,catena2007einstein,chiba2013conformal,chiba2008extended,
	qiu2012reconstruction,li2014generating} and here we shall just give a brief review. See\cite{capozziello2006cosmological,nojiri2006modified,briscese2007phantom,nojiri2011unified,wetterich2013universe,wetterich2014eternal,wetterich2014hot} for different points of view. We show here with an illustrative example\cite{deruelle2011conformal} how the observables do not depend on the frame one computes them. In particular, we consider an essential quantity in cosmology, the redshift.
	
	For simplicity's sake, let us assume that matter is minimally coupled to an expanding background. The line element is given by
	\begin{align}
		ds^2=-dt^2+a^2(t)d^2\sigma_{(K)},
	\end{align}
	where $a(t)$ is the scale factor, $d^2\sigma_{(K)}$ is an homogeneous and isotropic 3 dimensional space and $K=\pm 1,0$. The expansion of the universe is dictated by the Friedmann equation 
	\begin{align}
		\left(\frac{\dot a}{a}\right)^2\equiv H^2=\frac{8\pi G}{3}\rho-\frac{K}{a^2}\,,
	\end{align}
    and due to such expansion, one observes a cosmological redshift
	\begin{align}
		E_{obs}=\frac{E_{emit}}{1+z}\,.
	\end{align}
	This is regarded as a ``proof'' of expansion, or at least this is how we \textit{interpret} the data.
	
	However, as we are interested in conformal frames, let us go to a special frame by choosing $\Omega=1/a$ and working with conformal time $d\eta=dt/a$. By doing so, we go to a frame where the line element is given by
	\begin{align}
		d\tilde s^2=\Omega^2 ds^2=-d\eta^2+d\sigma^2_{(K)}\,,
	\end{align}
	which is a \textit{static universe} and, as such, photons do not redshift. Then, one may wonder whether this frame is unphysical. Let us show that it is as physical as any other.
	
	Recall that the mass of a fermion, e.g. an electron, is rescaled as \eqref{eq:fermion}
	\begin{align}\label{eq:mass}
		\tilde m=\Omega^{-1} m=\frac{m}{1+z}
	\end{align}
	where we used the usual definition of redshift, i.e. $a^{-1}=1+z$. Consequently, the Bohr radius scales as $\propto m^{-1}$ and, hence, the atomic energy levels scale inversely proportional, i.e. $\propto m$. Equation \eqref{eq:mass} yields time dependent energy levels, namely
	\begin{align}
		\tilde E_n=\frac{E_n}{1+z}\,,
	\end{align}
	where $\tilde E_n$ is the energy levels in the static frame and $E_n$ is the energy levels in the Jordan (matter) frame. Thus, it is clear that frequency of photons emitted at a time $z(\eta)$ from a level transition $n\to n'$ is
	\begin{align}
		\tilde E_{nn'}=\frac{E_{nn'}}{1+z}\,,
	\end{align}
	which is exactly what we observe as Hubble's law. Then, how do we interpret the Cosmic Microwave Background (CMB) photons in this static frame? First, the universe was in thermal equilibrium at $T=2.725K$ when the electron mass was more than $10^3$ times smaller, that is at a time $z>10^3$. Afterwards, CMB photons have never redshifted. How about the rate of scattering/interaction? The Thomson cross section $\tilde\sigma_T$ scales as $\propto m^{-2}$ and the electron density $\tilde n_e$ is indeed constant in the static frame. Relating them to the matter frame quantities we respectively have:
	\begin{align}
		&\tilde \sigma_{T}={\sigma_T}{(1+z)^2}&\quad &{\rm and}&\quad&\tilde n_e=\frac{n_e}{(1+z)^3}	\,.
	\end{align}	
	Therefore, the rate of scattering per unit of proper time remains unchanged, 
		\begin{align}
			\tilde \sigma_{T}\tilde n_e d\eta=\frac{\sigma_{T}\,n_e}{1+z}d\eta=\sigma_{T}\,n_e\,dt	\,.
		\end{align}	
	So far we showed that observed physics, e.g. redshift, is independent of the frame but interpretations of such observables most likely differ. In this way, it might be more appropriate to call them ``representations'' rather than frames\cite{deruelle2011conformal}. One last interesting consequence of the previous example is that the metric itself is non-observable.

	\section{Conformal frame ``dependence'' of inflation \label{conf}}
	In this section, let us put to good use all of the previously discussed. It is hopefully clear that there is no frame more physical than others and that observables do not depend on the frame one computes them. Bearing this in mind let us consider a two field inflationary model with the action given by
	\begin{align}
		S=\int d^4x \sqrt{-g}\left\{\frac{1}{2}R+{\cal L}_{inf}(\phi)\right\}+\sqrt{-\bar g}~{\cal L}_m(\chi)\,,
	\end{align}
	where $\phi$ and $\chi$ are scalar fields and $g$ and $\bar g$ are related by a conformal transformation, i.e.
	\begin{align}\label{eq:conf2}
		\bar g_{\mu\nu}=F^{-1}(\phi)g_{\mu\nu}\,.
	\end{align}
	This action is not neither in the Jordan frame nor Einstein frame form, but emphasises the idea of our model. Basically, we assume that $\phi$ drives inflation and $\chi$ is an spectator matter field, so-called the curvaton\cite{moroi2002cosmic,enqvist2002adiabatic,lyth2002generating}. Thus, the curvaton does not contribute to the inflationary dynamics and need not be in an accelerated expanding universe, i.e. metric $g$, but feels the metric $\bar g$, which might not be inflating. The same description applies for matter fields universally coupled to the metric $\bar g$. Since we define inflation as an accelerated expansion, we refer to the fact that matter might be in a very different universe as \textit{conformal ``dependence'' of inflation}\cite{Domenech:2015qoa}.

	In order to avoid any confusion
	the action in the Jordan (matter) frame is given by
		\begin{align}
			S=\int d^4x \sqrt{-\bar g}\left\{\frac{1}{2}F(\phi)\bar R+\bar{\cal L}_{inf}(\phi)-\frac{1}{2}\bar g^{\mu\nu}\nabla_\mu\chi\nabla_\nu\chi-\frac{1}{2}\bar m^2\chi^2 \right\}\,,
		\end{align}
	while the Einstein frame counterpart is given by
			\begin{align}
				S=\int d^4x \sqrt{-g}\left\{\frac{1}{2}R+{\cal L}_{inf}(\phi)-\frac{1}{2 F(\phi)} g^{\mu\nu}\nabla_\mu\chi\nabla_\nu\chi-\frac{1}{2}m^2\chi^2\right\}\,,
			\end{align}
	where we redefined $m\equiv F^{-1/2}\bar m$ and ${\cal L}_{inf}(\phi)$ is the Einstein frame counterpart of $\bar{\cal L}_{inf}(\phi)$, as in general the functional form changes. Notice that in the Einstein frame there is an explicit coupling between inflaton and curvaton while in the Jordan frame the curvaton is minimally coupled to the metric $\bar g$. Thus, the ``natural'' frame of the curvaton is the Jordan frame while the ``natural'' frame of inflation, understood as an accelerated expansion, is the Einstein frame.
	
	At this point, note is in order. We did not specify the functional form of $F(\phi)$ or, in other words, the matter metric $\bar g$. A particular choice of $F(\phi)$ corresponds to fixing the model. Nevertheless, this is not an obstacle as we assume inflation in the Einstein frame and the curvaton is assumed to be subdominant. Thus, the exact form of $F(\phi)$ is not relevant for the dynamics of inflation. The main point is that if the field $\chi$ contributes to the total curvature power spectrum, as it is the case for the curvaton model, then $F(\phi)$ may become extremely important. It leaves an observable imprint of the matter point of view.
	
	For the sake of simplicity, we considered a simple inflationary model, so-called power-law inflation\cite{lucchin1985power}, as an illustrative example. We proceed as follows: i) we compute the curvature power spectrum due to the inflaton in the Einstein frame while completely neglecting the curvaton, ii) we investigate how does $\bar g$ behaves for some choice of $F(\phi)$ and iii) we compute the power spectrum of the curvaton in the Jordan frame in some interesting examples.	
	
	\paragraph{Brief review of power-law (Einstein frame):}
	The inflaton lagrangian in the power-law model \cite{lucchin1985power} is given by
	\begin{align}
	{\cal L}_{inf}(\phi)=-\frac{1}{2} g^{\mu\nu}\nabla_\mu\phi\nabla_\nu\phi-V_0 {\rm e}^{-\lambda\phi}.
	\end{align}
	An exact solution to the equations of motion yields
	\begin{align}
	&a=a_0\left(\frac{t}{t_0}\right)^p \quad & H\equiv\frac{\dot a}{a}=\frac{p}{t}\\
	&\phi=\frac{2}{\lambda}\ln\left(\frac{t}{t_0}\right) \quad & (0<t<\infty)
	\end{align}
	where $p=2/\lambda^2$ and $\lambda^2 V_0 t_0^2=2(3p-1)$. Slow roll inflation occurs for $p\gg1$. The curvature and tensor power spectrum at horizon crossing ($k=aH$) are found to be\cite{kodama1984cosmological,mukhanov1992theory,lucchin1985power,lyth1992curvature}, respectively,
	\begin{align}
	&{\cal P}_{{\cal R}_c}=\frac{p}{8\pi^2}\frac{H^2}{M_{pl}^2}=\frac{p}{8\pi^2}\frac{H_0^2}{M_{pl}^2}\left(\frac{k}{k_0}\right)^{\frac{-2}{p-1}}\,,\quad&{\cal P}_{{\cal T}}=\frac{2}{\pi^2}\frac{H^2}{M_{pl}^2}\,,
	\end{align}
	where we recovered the units and from which one can extract the spectral index, that is
	\begin{align}\label{eq:index}
	n^\phi_s-1=\frac{d\ln{\cal P}_{{\cal R}_c}}{d\ln k}=\frac{-2}{p-1}=n_{\cal T}\,,
	\end{align}
	where the superscript $\phi$ emphasise that it is the contribution from the inflaton. If there is no other fields contributing to the power spectrum, the former result is observable through the CMB. Let us move on to the curvaton.
	
	\subsection{Matter point of view (Jordan frame)}
\begin{figure}[t]
				\centering
				\includegraphics[width=0.45\columnwidth]{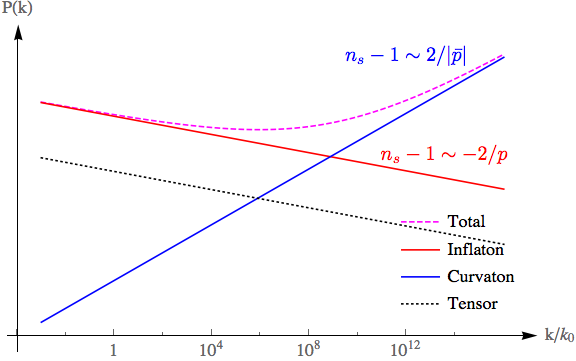}\hspace{0.5cm}
				\includegraphics[width=0.45\columnwidth]{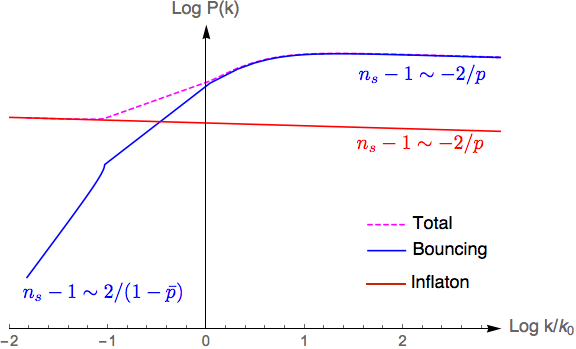}
				\caption{Total power spectrum. Left: Jordan power law curvaton. In red the inflaton curvature perturbation, in blue the one from the curvaton and in magenta the sum. A black line shows the tensor power spectrum. Right: Jordan bounce curvaton. In red inflaton curvature perturbation and in blue the contribution from the curvaton with a initial contracting phase.  Lastly, the magenta dashed line is the total power spectrum in the bouncing case.}
				\label{fig:1}
			\end{figure}

	We choose two different couplings to the curvaton. The first one leads us to another power law but with a different power law index and the second one yields a bouncing matter metric.
	\paragraph{Jordan frame power-law.}
	Let us choose that the curvaton metric is related to the inflationary metric by
	\begin{align}
		F(\phi)
		=\textrm{e}^{\gamma\lambda\phi}=(t/t_0)^{2\gamma}\,,\label{fcase1}
	\end{align}
	where $\gamma$ is a free parameter. Plugging the former to equation \eqref{eq:conf2} yields
		\begin{align}
		&\bar a={a}_0\left(\frac{\bar t}{{\bar t}_0}\right)^{\bar p}\,, \quad & \bar H\equiv\frac{1}{\bar a}\frac{d\bar a}{d\bar t}=\frac{\bar p}{\bar t}\,,
		\end{align}
		where
		\begin{align}
		&\bar p>1~(0<\bar t<\infty)\,, \quad& \bar p<1~ (-\infty<\bar t<0)
		\end{align}
		and
		\begin{align}
		&\bar p-1=\frac{p-1}{1-\gamma}\,,\quad &\bar t_0=\frac{t_0}{1-\gamma}\,,
		\end{align}
		which is another power law type universe. Interestingly, although there is an accelerated expansion in the Einstein frame, the curvaton might feel a different evolution, depending on the choice of $\gamma$. Concretely, for $\bar p>1$ ($\gamma <1$) we have inflation, $0<\bar p<1$ ($\gamma >p$) leads to a decelerated contracting universe and for $\bar p<0$ ($1<\gamma <p$) it is a superinflationary universe. The latter is the most interesting case as it yields a blue tilted curvaton power spectrum, i.e. 
		\begin{eqnarray}
			\tilde{n}_\chi-1=\frac{-2}{\tilde{p}-1}>0 &\qquad (\bar p <0)\,,
		\end{eqnarray}
		where we assumed that the curvaton instantly reheats the universe\cite{moroi2002cosmic,enqvist2002adiabatic,lyth2002generating}. It has exactly the same form as the inflaton power spectrum \eqref{eq:index} but with $\bar p$ instead of $p$, as one expects from a power law universe. The total power spectrum is shown in left figure \ref{fig:1}.
		It should be noted that once we fix $\gamma$ we are fix the curvaton Jordan frame, which obviously leads to different models, different observational results.
		
	\paragraph{Jordan frame bounce.}
	Finally we give another example, a bouncing Jordan metric, where this time
	\begin{align}
		F(\phi)=\Big(1+\textrm{e}^{{-\gamma\lambda}\phi/2}\Big)^{-2}
		=\Big(1+(t/t_0)^{-\gamma}\Big)^{-2}\,.
		\label{fcase2}
	\end{align}
	Again, plugging this form into equation \eqref{eq:conf2} roughly yields (for $\bar p<1$)
	\begin{align}
		\bar{a} \approx \left\{
		\begin{array}{ccc}
			a_0(-\bar{t}/\bar{t}_0)^{\bar{p}} & \quad |\bar{t}|\gg\bar{t_0}
			& \quad(\bar{t}<0)
			\\
			a_0(\bar{t}/\bar{t}_0)^{p} & \quad \bar{t}\gg\bar{t_0}
		\end{array}
		\right.\,.
	\end{align}
	Initially the curvaton is in a contracting universe which bounces and catches up with the Einstein frame power law. The total power spectrum is shown in right figure \ref{fig:1}. The main feature in this case is a blue tilt at long scales, which if the curvaton dominates at short scales leads to a suppression of the power spectrum for large $k$.

	\section{Discussion and Conclusions\label{matter}}
	
	We argued that in higher dimensional theories the presence of dilatonic scalar fields, i.e. that non-minimally couple to gravity and/or matter, is usually expected. When such a non-minimal coupling is present, the notion of conformal frames (or ``representations'') appears, in particular the Jordan (matter) frame and the Einstein frame. We reviewed the physical equivalence of observables between frames by showing an illustrative example where the redshift instead of being interpreted as ``proof'' of expansion is due to a time dependent electron mass.
	
	We then considered a two field model, where the second field is an spectator field, a curvaton, and it is minimally coupled to a metric which is not inflating. We define inflation in the Einstein frame as an accelerated expansion and we call the fact that the curvaton need not be in such a metric \textit{conformal ``dependence'' of inflation}. Thus, we easily get a blue tilt contribution from the curvaton to the curvature power spectrum by minimally coupling the curvaton to a superinflationary universe. Furthermore, we considered another example where the curvaton feels a bouncing universe.
	
	In this way, we emphasized the \textit{physical equivalence} between conformal frames in cosmology while paying special attention to how \textit{matter couples} to gravity, which might be observationally relevant if for example the curvaton mechanism is present. Thus, without specifying the matter frame a model is not completely defined. Moreover, the physical equivalence between frames related by a generalized transformation, called disformal transformation\cite{Bekenstein:1992pj}, has also been shown\cite{minamitsuji2014disformal,bettoni2013disformal,deruelle2014disformal,Watanabe:2015uqa,Motohashi:2015pra,tsujikawa2014disformal,Domenech:2015hka,Domenech:2015tca}.
	
    As a final remark, let us consider the importance of the matter coupling in a more general theory, such as Horndeski\cite{horndeski1974second,kobayashi2011generalized} and its extensions\cite{gleyzes2014exploring,Gleyzes:2014dya,Gao:2014soa,Deffayet:2011gz}. Horndeski model implicitly assumes that matter is minimally coupled, namely
	\begin{align}
		S=\int d^4x \sqrt{-g}\left\{{\cal L}_{H}(\phi,g,R)+{\cal L}_{matter}(\psi,A)\right\}\,,
	\end{align}     
	where ${\cal L}_{H}$ is a general function of a non-minimally coupled scalar field $\phi$ which yields second order differential equations of motion. On the other hand, if the matter sector is minimally coupled to a different metric, i.e.
	\begin{align}
		S=\int d^4x \sqrt{-g} {\cal L}_H(\phi,g,R)+\sqrt{-\bar g}\,{\cal L}_{matter}(\psi,A)\,,
	\end{align}  
	we clearly have different model which indeed yields different features. In particular, if $\bar g$ is related to $g$ by a derivative dependent disformal transformation, the latter theory falls in the category of beyond Horndeski theory and may experience a breaking of the Vainshtein mechanism inside astrophysical bodies\cite{Kobayashi:2014ida,Koyama:2015oma,Saito:2015fza}. For these reasons, further study on how matter could couple to gravity might be an interesting direction. 
	
	\section*{Acknowledgments}
		GD would like to thank J. Fedrow, J.O. Gong, G. Leung, R. Namba, T. Qiu and J. White for useful discussions on the conformal ``dependence'' of inflation and N. Deruelle and R. Saito for interesting discussions on matter couplings in Horndeski. This work was supported in part by MEXT KAKENHI Grant Number 15H05888.
	
	\bibliographystyle{ws-procs961x669}
	\bibliography{bibliography}
		
\end{document}